\documentclass[%
 reprint,
 amsmath,amssymb,
 aps,
]{revtex4-2}

\usepackage{graphicx}% Include figure files
\usepackage{subcaption}
\usepackage[percent]{overpic}

\usepackage{dcolumn}% Align table columns on decimal point
\usepackage{bm}% bold math
\usepackage{xcolor}
\usepackage{mathtools}
\usepackage{amsmath}
\usepackage{url}
\usepackage{lipsum}
\usepackage{amsmath,amssymb,amsthm} 

\usepackage{epigraph} 
\usepackage{hyperref}

\begin{document}

\preprint{APS/123-QED}

\title{\textbf{Temporal Interfaces in Metamaterials with Shape-Varying Inclusions}}

\author{Ishtiaque Rahman}
\email{ishtiaque.rahman@aalto.fi}
\author{Mohamed Hesham Mostafa}%
\author{Viktar Asadchy}%
 \email{viktar.asadchy@aalto.fi}
\affiliation{%
Department of Electronics and Nanoengineering, Aalto University, P.O. Box 15500, FI-00076 Aalto, Finland\\
\\
}%

\begin{abstract}
Temporal modulation of metamaterials is commonly achieved by varying constituent material properties, while changes in meta-atom geometry offer an additional means of controlling their effective response. Here, we develop an analytical and numerical framework for temporal scattering in metamaterials composed of cylindrical inclusions undergoing abrupt geometric transformations. Changing the radius of circular cylinders modifies the filling fraction and scalar in-plane effective permittivity, enabling temporal reflection, field amplification or attenuation, and frequency conversion. An area-preserving transformation from circular to elliptical cross sections introduces anisotropy, producing angle-dependent temporal scattering and redirecting the energy flow while conserving the wavevector. We derive these effects using effective-medium theory and temporal boundary conditions and compare the predictions with full-wave time-domain simulations. The simulations reproduce the main scattering and temporal-aiming trends, with larger angular deflections than predicted by the quasistatic model at higher permittivity contrasts. This suggests that moving toward the resonant regime may offer opportunities for stronger temporal aiming. These results identify inclusion geometry as a means of controlling wave amplitude, frequency, and energy-flow direction at temporal interfaces.
\end{abstract}

\maketitle

\section{Introduction}

The study of electromagnetic wave interactions with structured materials has undergone a fundamental shift driven by growing interest in temporal and spatiotemporal modulation of material properties \cite{galiffi2022photonics,Asgari:24, mostafa2024temporal, Boltasseva2024, Hayran2023, OrtegaGomez2023, Taravati2022, Caloz2020PartII}. Conventionally, electromagnetic responses in bulk media were engineered through spatial variation of material parameters, as in metamaterials, photonic crystals, and periodic composites \cite{PhysRevLett.58.2486, PhysRevLett.58.2059, joannopoulos2008photonic, Soukoulis2011, Yu2014, Zheludev2012, Glybovski2016}. Such spatially modulated systems have enabled remarkable phenomena, including negative refraction and near-zero permittivity \cite{cui2009metamaterials,soukoulis2011past}. However, the introduction of time as an additional modulation dimension has opened a fundamentally new regime of wave-matter interaction with no spatial analogue.

Central to temporal modulation of electromagnetic media is the concept of a temporal interface, at which material parameters are switched uniformly throughout space at a single instant in time~\cite{morgenthaler1958velocity, Mirmoosa2024, PachecoPena2025, Galiffi2025, Shlivinski2018}. Unlike spatial interfaces, temporal interfaces conserve the spatial momentum of the electromagnetic wave while allowing energy exchange with the modulated medium, leading to wave amplification or attenuation. The study of temporal interfaces has given rise to a range of unique phenomena such as temporal reflection and wave amplification, which has driven a surge of interest in this research direction \cite{galiffi2022photonics}.

A particularly compelling recent development is the extension of temporal-interface physics to metamaterials and metasurfaces assembled from resonant meta-atoms whose constituent properties are varied in time \cite{shi2016dynamic,lee2018linear,shcherbakov2019photon,pachecopena2021spatiotemporal,garg2022modeling,wang2023metasurface,garg2024homogenization,duan2024linear,prudencio2024engineering,wang2025expanding,rawat2026generation}.
In such systems, temporal scattering is governed not only by the instantaneous constitutive change but also by the modal spectrum and dispersion of the inclusions. Material and structural resonances can therefore strongly enhance the response to a modest microscopic material modulation. 
In all of these approaches, however, the physical geometry of the inclusions remains fixed, while their permittivity, conductivity, loss, resonance frequency, or inter-resonator coupling is modulated in time. An equally compelling, yet largely unexplored, alternative is to modulate the geometry of the inclusions themselves. An early study~\cite{panagiotidis2022inelastic} investigated light scattering from a single dielectric sphere with a periodically time-varying radius. However, it did not address the collective effective-medium response of an array of such shape-varying inclusions.
Such geometry-driven modulation provides an additional degree of freedom for controlling both the magnitude and anisotropy of the effective permittivity.

In this paper, we develop a theoretical description of temporally shape-varying metamaterials and compare its predictions with full-wave time-domain simulations. We consider two transformations of cylindrical inclusions: (i) an abrupt radius change, which modifies the filling fraction and scalar in-plane effective permittivity, and (ii) an area-preserving change from circular to elliptical cross sections, which introduces in-plane anisotropy. We assume lossless, nondispersive constituent materials and evaluate the effective permittivities in the quasistatic limit. The full-wave simulations additionally capture the structural dispersion associated with the finite inclusion size and lattice period. The first configuration enables temporal reflection, field amplification or attenuation, and frequency conversion. The second produces angle-dependent temporal scattering and temporal aiming~\cite{pacheco2020temporal}, redirecting the energy flow while conserving the wavevector. Together, these configurations demonstrate how changes in inclusion geometry can control the temporal scattering response and the direction of energy transport.
One possible physical platform for realizing shape-varying metamaterials is a patterned plasma discharge. Experiments have shown that plasma filaments with cylindrical shapes generated by dielectric-barrier discharges can self-organize into periodic arrays whose unit-cell configuration,  dimensions of scattering elements, and period can be reconfigured by varying the discharge conditions~\cite{fan2019spatiotemporally,wang2019woodpile,
gao2020structural,fan2025tailored,barbuto2025modeling}. Moreover, femtosecond-scale laser ionization has been recently demonstrated, suggesting a possible route toward sufficiently rapid plasma creation for temporal-interface operation at microwave and THz frequencies~\cite{gao2013femtosecond,huang2025terahertz}. 
More recently, a closely related concept for realizing anisotropic temporal interfaces using stratified metamaterials was proposed in \cite{naylor2026anisotropic}. However, achieving rapid temporal modulation of the large planar layers considered in that study may present significant practical challenges. On the contrary, our proposed cylindrical geometry is naturally compatible with reconfigurable plasma-column arrays.

\raggedbottom

\section{Geometry of the problem and definitions}
\label{sec:geometry}

\begin{figure*}[t]
    \centering
    \includegraphics[width=\linewidth]{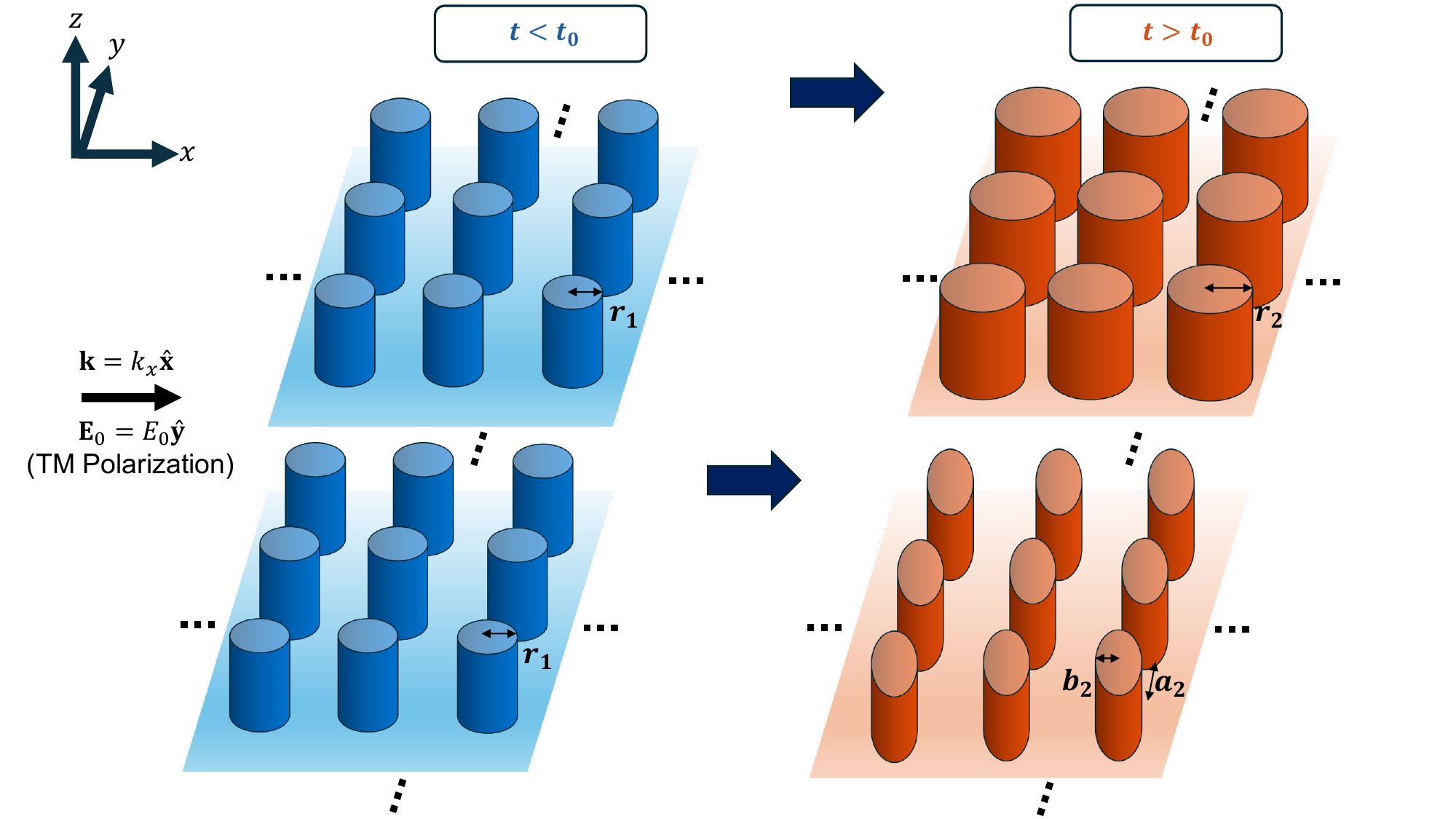}

    \caption{%
        Geometry of the temporally shape-varying metamaterials
        considered in this work. The cylindrical inclusions are
        oriented along the $z$-axis and arranged periodically
        in the transverse $xy$-plane. In the first case (top), we have the Isotropic-to-isotropic temporal transition in which
        circular cylinders abruptly change their radius from
        $r_1$ to $r_2$, thereby modifying the filling fraction
        and the scalar in-plane effective permittivity. The second scenario (bottom) depicts an Isotropic-to-anisotropic temporal transition in which
        circular cylinders of radius $r_1$ abruptly transform
        into elliptical cylinders with semi-axes $a_2$ and $b_2$.
    }
    \label{fig:geometry}
\end{figure*}

We consider two-dimensional metamaterial geometries formed by periodic arrays of effectively infinite cylindrical inclusions oriented along the $z$-axis (in fact, it is sufficient that they are much longer than the operational wavelength), as illustrated in Fig.~\ref{fig:geometry}. This choice substantially simplifies the full-wave time-domain simulations while retaining the essential physics of geometry-driven temporal modulation. It is also experimentally attractive, since arrays of elongated inclusions are more practical to realize than arrays of full three-dimensional particles, particularly using plasma column arrays~\cite{fan2019spatiotemporally,wang2019woodpile,gao2020structural,fan2025tailored,barbuto2025modeling}. The corresponding theoretical treatment can be extended to three-dimensional inclusions, such as spheres and ellipsoids, by replacing the homogenization relations with their three-dimensional counterparts.

The cylinder inclusions with permittivity $\varepsilon_{\mathrm{i}}$ are embedded in a homogeneous host medium (environment) with permittivity $\varepsilon_{\mathrm{e}}$. Both materials are assumed to be isotropic, nonmagnetic, lossless, and dispersionless, so that the temporal modulation changes only the inclusion geometry while leaving the constituent material parameters unchanged. We define the permittivity contrast as $\varepsilon=\varepsilon_{\mathrm{i}}/\varepsilon_{\mathrm{e}}$ and consider both regimes $\varepsilon<1$ and $\varepsilon>1$ for completeness. Material dispersion and loss can be incorporated in future studies using established theoretical frameworks~\cite{mirmoosa2022dipole,ptitcyn2023floquet,ganfornina2026generalized}. For cylinders invariant along $z$, we can define the filling fraction as the transverse area fraction, $f=n_{\mathrm{2D}}A$, where $n_{\mathrm{2D}}$ is the number of cylinders per unit area in the $xy$-plane and $A$ is the cross-sectional area of one cylinder.

Two temporal transformations are considered in this work. In Sec.~\ref{isotropic}, the circular cylinders abruptly change their radii from $r_1$ to $r_2$ at the switching instant $t=t_0$, as shown in Fig.~\ref{fig:geometry}(a). Their cross-sectional area consequently changes from $\pi r_1^2$ to $\pi r_2^2$, modifying the filling fraction and hence the scalar in-plane effective permittivity. As the circular cross section is rotationally symmetric, the metamaterial is always isotropic in the $xy$-plane, with $\varepsilon_{\mathrm{eff},x}=\varepsilon_{\mathrm{eff},y}$.

In Sec.~\ref{anisotropic}, the initially circular cylinders abruptly transform into elliptical cylinders with semi-axes $a_2$ and $b_2$, as illustrated in Fig.~\ref{fig:geometry}(b). To isolate the effect of shape from that of inclusion volume, we impose cross-sectional-area conservation, $a_2b_2=r_1^2$, so that the filling fraction remains unchanged across the temporal interface. The unequal depolarization factors of the elliptical cross section then produce an anisotropic in-plane effective permittivity,

\begin{align}
\overline{\overline{\varepsilon}}_{\mathrm{eff},2}^{\,\parallel}
=
\begin{bmatrix}
\varepsilon_{\mathrm{eff},2x} & 0\\
0 & \varepsilon_{\mathrm{eff},2y}
\end{bmatrix},
\qquad
\varepsilon_{\mathrm{eff},2x}
\neq
\varepsilon_{\mathrm{eff},2y},
\label{eq1new}
\end{align}
thereby realizing an isotropic-to-anisotropic temporal transition solely through modulation of the inclusion shape.

\section{Effective-medium theory for the time-varying metamaterial}
\label{sec:effective_medium}

To describe temporal scattering from the shape-varying metamaterial, we adopt an effective-medium approximation. Specifically, we homogenize the metamaterial independently before and after the temporal jump, obtaining the effective permittivities $\overline{\overline{\varepsilon}}_{\mathrm{eff},1}$ and $\overline{\overline{\varepsilon}}_{\mathrm{eff},2}$, respectively. We then approximate the temporal reflection and transmission coefficients of the structured medium by those of a temporal interface in a homogeneous medium characterized by the same effective permittivities. This approximation is applicable in the long-wavelength regime~\cite[p.~151]{sihvola1999electromagnetic}, where both the transverse dimensions of the cylinders and the lattice period are much smaller than the wavelength in the effective medium before and after the temporal transition. As demonstrated below, this effective-medium approximation accurately captures the temporal scattering response, yielding results in close agreement with full-wave time-domain simulations in the long-wavelength regime. 

The effective permittivity is determined by the electric polarizability of the individual inclusions and therefore depends on their depolarization factors. For a general ellipsoid with semi-axes $a_x$, $a_y$, and $a_z$, the depolarization factor along the principal direction $j\in\{x,y,z\}$ is given by~\cite[p.~63]{sihvola1999electromagnetic}
\begin{align}
N_j =
\frac{a_xa_ya_z}{2}
\int_{0}^{\infty}
\frac{\mathrm{d}s}
{(s+a_j^2)
\sqrt{(s+a_x^2)(s+a_y^2)(s+a_z^2)}}.
\label{eq:depolarization_factor_ellipsoid}
\end{align}
Here, $s$ is an auxiliary integration variable with dimensions of length squared.
For an infinitely long cylinder oriented along the $z$-axis and having an elliptical cross section with semi-axes $a$ and $b$ along $x$ and $y$, respectively, these factors reduce to~\cite{sihvola2005metamaterials}
\begin{align}
N_x=\frac{b}{a+b},
\qquad
N_y=\frac{a}{a+b},
\qquad
N_z=0.
\label{eq:depolarization_factors_2d}
\end{align}
For the circular-cylinder limit, $a=b=r$, one obtains $N_x=N_y=1/2$, corresponding to an isotropic response in the transverse $xy$-plane.

For aligned cylindrical inclusions in an isotropic host, the generalized Maxwell Garnett mixing rule gives the effective-permittivity component along the principal direction $j$~\cite[p.~67]{sihvola1999electromagnetic}. Using the normalized permittivity contrast $\varepsilon=\varepsilon_{\mathrm{i}}/\varepsilon_{\mathrm{e}}$, it can be written as
\begin{align}
\frac{\varepsilon_{\mathrm{eff},j}}
     {\varepsilon_{\mathrm{e}}}
=
1+
\frac{f(\varepsilon-1)}
     {1+(1-f)N_j(\varepsilon-1)}.
\label{eq:MG_elliptical_cylinders}
\end{align}
Here, $f=n_{\mathrm{2D}}A$ is the filling fraction, where $A=\pi ab$ is the inclusion cross-sectional area. For a circular cylinder, Eq.~\eqref{eq:MG_elliptical_cylinders} becomes independent of the in-plane direction and reduces to
\begin{align}
\frac{\varepsilon_{\mathrm{eff}}^{\,\parallel}}
     {\varepsilon_{\mathrm{e}}}
=
1+
\frac{2f(\varepsilon-1)}
     {2+(1-f)(\varepsilon-1)}.
\label{eq:MG_circular_cylinders}
\end{align}

For the isotropic-to-isotropic transition considered in Sec.~\ref{isotropic}, the depolarization factors remain fixed at $N_x=N_y=1/2$, while the radius change from $r_1$ to $r_2=c r_1$ modifies the filling fraction from $f_1=n_{\mathrm{2D}}\pi r_1^2$ to $f_2=n_{\mathrm{2D}}\pi r_2^2=c^2f_1$. Substitution of $f_1$ and $f_2$ into Eq.~\eqref{eq:MG_circular_cylinders} gives the scalar effective permittivities before and after the temporal interface.

For the isotropic-to-anisotropic transition considered in Sec.~\ref{anisotropic}, the cross-sectional area and filling fraction are conserved, but the circular cylinders transform into elliptical cylinders with semi-axes $a_2$ and $b_2$. The depolarization factors therefore change from $N_{1x}=N_{1y}=1/2$ to $N_{2x}=b_2/(a_2+b_2)$ and $N_{2y}=a_2/(a_2+b_2)$. Substitution of these factors into Eq.~\eqref{eq:MG_elliptical_cylinders} yields $\varepsilon_{\mathrm{eff},2x}\neq\varepsilon_{\mathrm{eff},2y}$ and hence the anisotropic in-plane effective-permittivity tensor defined in Eq.~\eqref{eq1new}. Thus, the two transformations considered here modify the effective response through complementary geometric mechanisms: the radius modulation changes the filling fraction while preserving rotational symmetry, whereas the circular-to-elliptical transformation changes the depolarization factors and breaks the in-plane rotational symmetry at a fixed filling fraction.

\section{Isotropic-to-Isotropic Temporal Interface}\label{isotropic}

Consider a monochromatic TM-polarized plane wave propagating along the $x$-axis, with its electric field along $y$ and magnetic field along $z$. Before and after the temporal interface at $t=t_0$, the metamaterial is isotropic in the transverse $xy$-plane, with effective permittivities $\varepsilon^{\,\parallel}_{\mathrm{eff,1}}$ and $\varepsilon^{\,\parallel}_{\mathrm{eff,2}}$, respectively. Throughout this section, we omit the superscript $\parallel$ for brevity. Both effective permittivities are assumed real, positive, and frequency independent, while the permeability remains equal to $\mu_0$.

The temporal interface is produced by abruptly changing the cylinder radius from $r_1$ to $r_2=c r_1$, where $c>0$ is a dimensionless scaling parameter. Since the lattice period remains unchanged, the filling fraction changes according to $f_2=c^2f_1$. For nonoverlapping circular cylinders arranged on a square lattice, both filling fractions must satisfy $f_{1,2}\leq f_{\mathrm{max}}=\pi/4\approx0.785$, which imposes the bound $c\leq\sqrt{f_{\mathrm{max}}/f_1}$. In the parameter maps below, we adopt the slightly lower plotting cutoff $f_{1,2}\leq0.75$.

Within the homogeneous effective-medium description, the modulation is spatially uniform and therefore conserves the wavevector $\mathbf{k}=k\hat{\mathbf{x}}$. The change in effective permittivity nevertheless changes the phase velocity and hence the wave frequency. The positive angular frequencies before and after the interface satisfy~\cite{morgenthaler1958velocity}
\begin{align}
    \frac{\omega_2}{\omega_1}
    = \frac{v_2}{v_1}
    = \sqrt{\frac{\varepsilon_{\mathrm{eff,1}}}
                  {\varepsilon_{\mathrm{eff,2}}}},
    \label{eq:freq_ratio}
\end{align}
where $v_j=1/\sqrt{\mu_0\varepsilon_{\mathrm{eff},j}}$ for $j=1,2$, with $\varepsilon_{\mathrm{eff},j}$ denoting absolute permittivity. At the conserved wavevector, the transmitted and reflected branches correspond to frequencies $+\omega_2$ and $-\omega_2$, respectively, and carry energy along $+x$ and $-x$.

In the absence of impulsive sources at the switching instant, the electric displacement $\mathbf{D}$ and magnetic flux density $\mathbf{B}$ remain continuous across the temporal interface. Applying these boundary conditions gives the temporal transmission and reflection coefficients~\cite{morgenthaler1958velocity,mostafa2024temporal}
\begin{align}
    T &=
    \frac{1}{2}
    \sqrt{\frac{\varepsilon_{\mathrm{eff,1}}}
               {\varepsilon_{\mathrm{eff,2}}}}
    \left(
    \sqrt{\frac{\varepsilon_{\mathrm{eff,1}}}
               {\varepsilon_{\mathrm{eff,2}}}}+1
    \right),
    \label{T_iso_meta}\\
    R &=
    \frac{1}{2}
    \sqrt{\frac{\varepsilon_{\mathrm{eff,1}}}
               {\varepsilon_{\mathrm{eff,2}}}}
    \left(
    \sqrt{\frac{\varepsilon_{\mathrm{eff,1}}}
               {\varepsilon_{\mathrm{eff,2}}}}-1
    \right).
    \label{R_iso_meta}
\end{align}
Here, $T$ and $R$ are the transmitted and reflected electric-field amplitudes normalized to the incident amplitude, with phases referenced to $t=t_0$. Under the stated assumptions, both coefficients are real and depend only on the effective-permittivity ratio.

To quantify the energy exchanged with the modulation, let $u_1$ denote the cycle-averaged incident energy density and $u_2=u_{\mathrm{T}}+u_{\mathrm{R}}$ the sum of the cycle-averaged energy densities of the transmitted and reflected waves. Their ratio is
\begin{align}
    \frac{u_2}{u_1}
    &=
    \frac{\varepsilon_{\mathrm{eff,2}}}
         {\varepsilon_{\mathrm{eff,1}}}
    \left(|T|^2+|R|^2\right)
    \notag\\
    &=
    \frac{1}{2}
    \left(
    \frac{\varepsilon_{\mathrm{eff,1}}}
         {\varepsilon_{\mathrm{eff,2}}}+1
    \right).
    \label{eq:energy_ratio_iso}
\end{align}
Thus, decreasing the effective permittivity increases the total wave energy, whereas increasing it reduces the total wave energy. These changes correspond to energy supplied by or transferred to the external modulation, despite the constituent materials being lossless.

For inclusions with a lower permittivity than the host, $\varepsilon=\varepsilon_{\mathrm{i}}/\varepsilon_{\mathrm{e}}<1$, increasing the radius ($c>1$) lowers the effective permittivity and produces amplification. For $\varepsilon>1$, the same radius increase raises the effective permittivity and produces attenuation. Decreasing the radius reverses these trends.

\begin{figure*}[t]
    \centering

    \begin{subfigure}{0.48\textwidth}
        \includegraphics[width=0.8\linewidth]{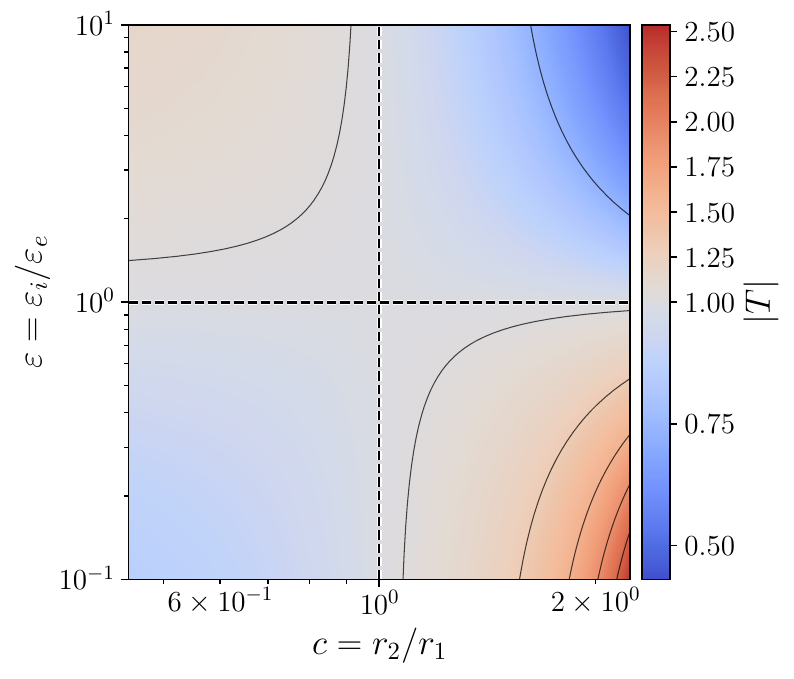}
        \caption{}
    \end{subfigure}
    % \hfill
    \begin{subfigure}{0.48\textwidth}
        \includegraphics[width=0.8\linewidth]{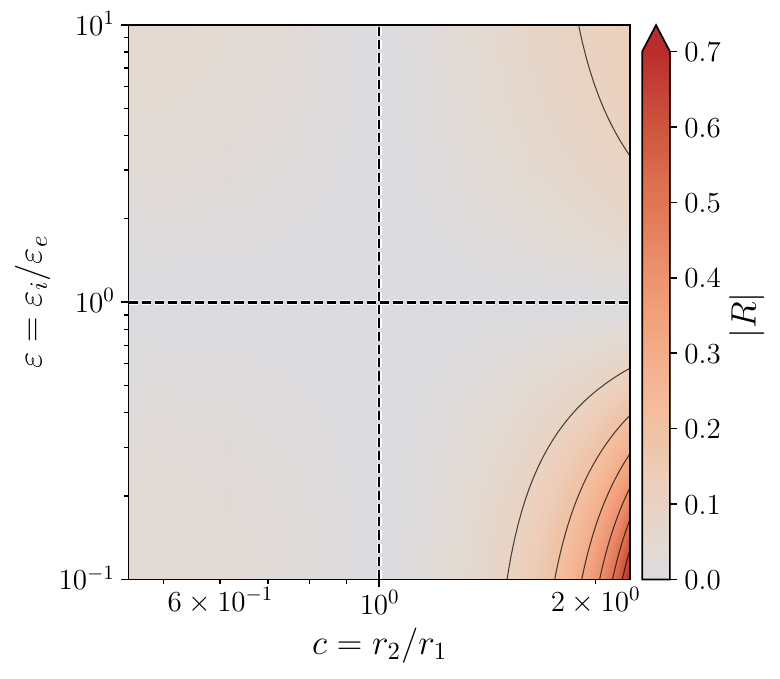}
        \caption{}
    \end{subfigure}
    
    \begin{subfigure}{0.48\textwidth}
        \includegraphics[width=0.8\linewidth]{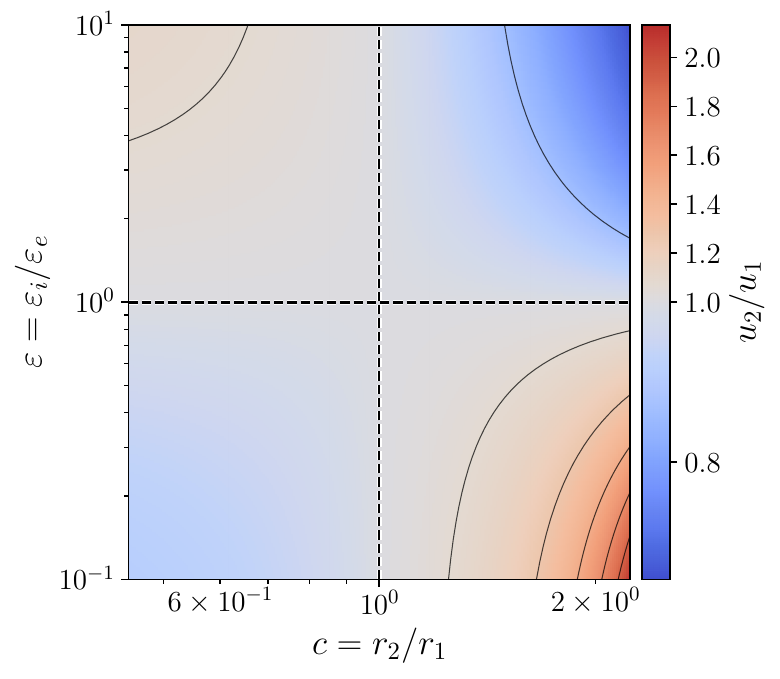}
        \caption{}
    \end{subfigure}
    % \hfill
    \begin{subfigure}{0.48\textwidth}
        \includegraphics[width=0.8\linewidth]{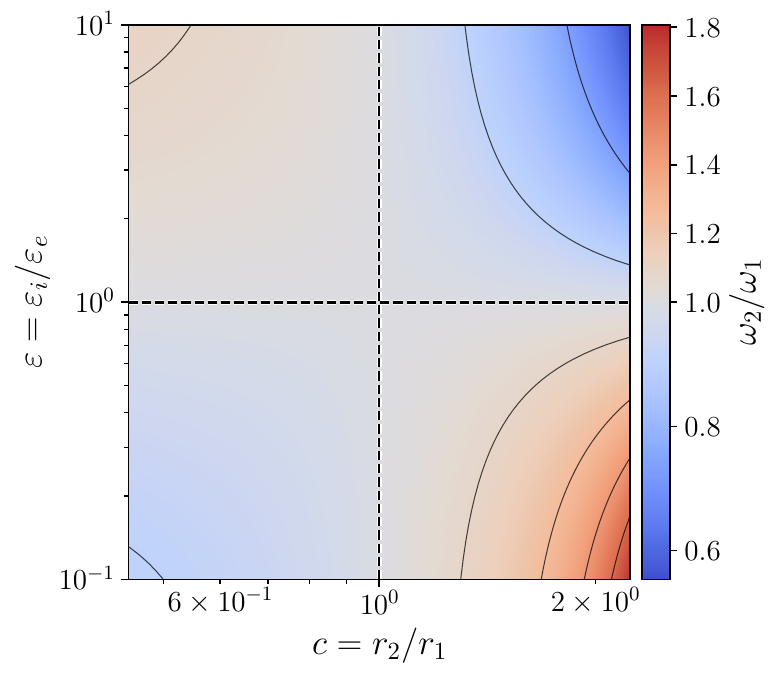}
        \caption{}
    \end{subfigure}
\caption{Temporal scattering at an isotropic-to-isotropic interface as a function of the inclusion radius ratio $c=r_2/r_1$ and permittivity contrast $\varepsilon=\varepsilon_{\mathrm{i}}/\varepsilon_{\mathrm{e}}$, with $f_1=0.15$. (a) Transmission amplitude $|T|$. (b) Reflection amplitude $|R|$. (c) Total wave-energy-density ratio $u_2/u_1$. (d) Frequency ratio $\omega_2/\omega_1$. Dashed lines mark $c=1$ and $\varepsilon=1$, where no temporal interface is produced.}
    \label{fig:pseudocolorplot_isotropic}
\end{figure*}

Figure~\ref{fig:pseudocolorplot_isotropic} plots the quantities in (\ref{eq:freq_ratio})--(\ref{eq:energy_ratio_iso}) as a function of the inclusion size ratio $c=r_2/r_1$ and permittivity contrast $\varepsilon=\varepsilon_i/\varepsilon_e$, with the initial filling fraction fixed at $f_1=0.15$. The dashed lines mark $c=1$ (unchanged inclusion size) and $\varepsilon=1$ (identical inclusion and host permittivities). Neither condition produces a temporal interface, giving $|T|=1$, $|R|=0$, $u_2/u_1=1$, and $\omega_2/\omega_1=1$.
As seen from Fig.~\ref{fig:pseudocolorplot_isotropic}~(a), the transmitted field is amplified whenever the geometric transformation lowers the effective permittivity, i.e., when expanding low-permittivity inclusions ($c>1$, $\varepsilon<1$) or contracting high-permittivity inclusions ($c<1$, $\varepsilon>1$). The other two combinations produce attenuation. 
The reflected-field magnitude in Fig.~\ref{fig:pseudocolorplot_isotropic}(b) reaches its largest values in the lower-right corner of the map and remains relatively weak elsewhere (with the current choice of $f_1$).

Figure~\ref{fig:pseudocolorplot_isotropic}(c) shows the combined energy density of the transmitted and reflected waves relative to that of the incident wave. Its amplification and attenuation regions coincide with those of the transmitted-field amplitude, confirming that the modulation supplies energy to the waves when the effective permittivity decreases and extracts energy when it increases. Figure~\ref{fig:pseudocolorplot_isotropic}(d) shows the accompanying frequency conversion. Conservation of the wavevector makes a decrease in effective permittivity produce a frequency upshift, whereas an increase produces a downshift. Energy amplification therefore accompanies frequency upshifting, consistently with
$u_2/u_1=[1+(\omega_2/\omega_1)^2]/2$.

A common feature of all four panels is the stronger response to inclusion expansion than to contraction over the plotted range. This asymmetry follows from $f_2=c^2f_1$: expansion can increase the filling fraction from $0.15$ to $0.75$, whereas contraction can remove at most the initial filling fraction of $0.15$. At the lower-right corner, $\varepsilon=0.1$ and $c=\sqrt{5}\approx2.24$, the model predicts $|T|\approx2.53$, $|R|\approx0.73$, $u_2/u_1\approx2.13$, and $\omega_2/\omega_1\approx1.81$, providing  the strongest temporal amplification, reflection, and frequency upconversion.

\begin{figure*}[t]
    \centering

    \begin{subfigure}{\textwidth}
        \begin{overpic}[width=\linewidth]{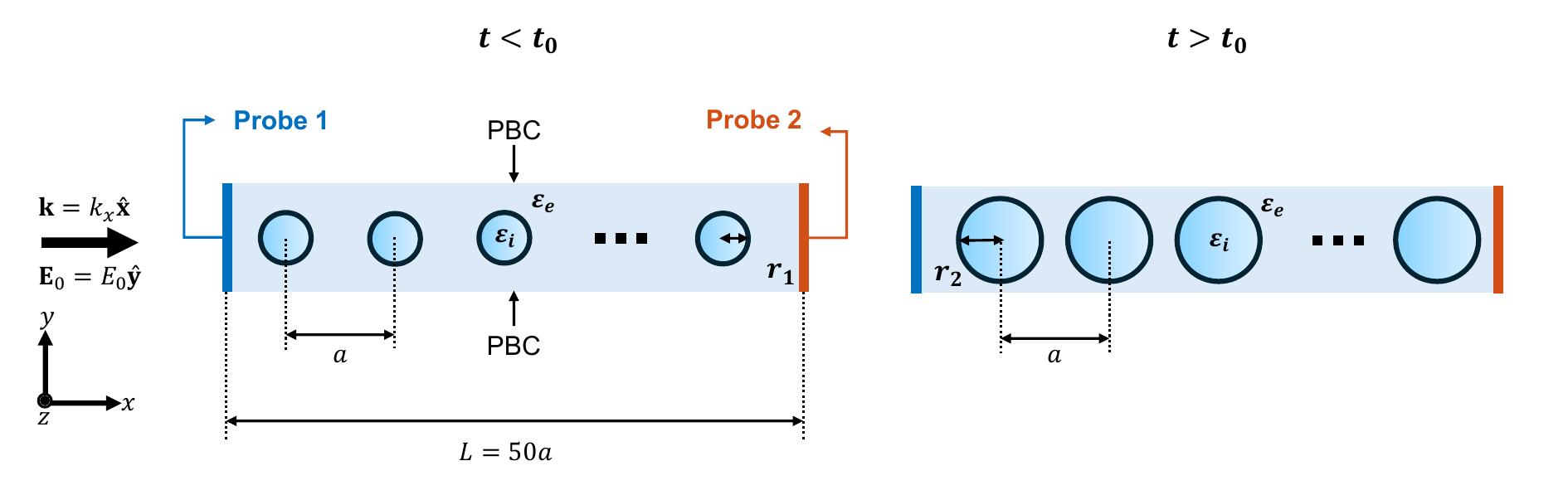}
            \put(0,28){(a)}
        \end{overpic}
        
    \end{subfigure}
    
    \begin{subfigure}{0.48\textwidth}
        \begin{overpic}[width=\linewidth]{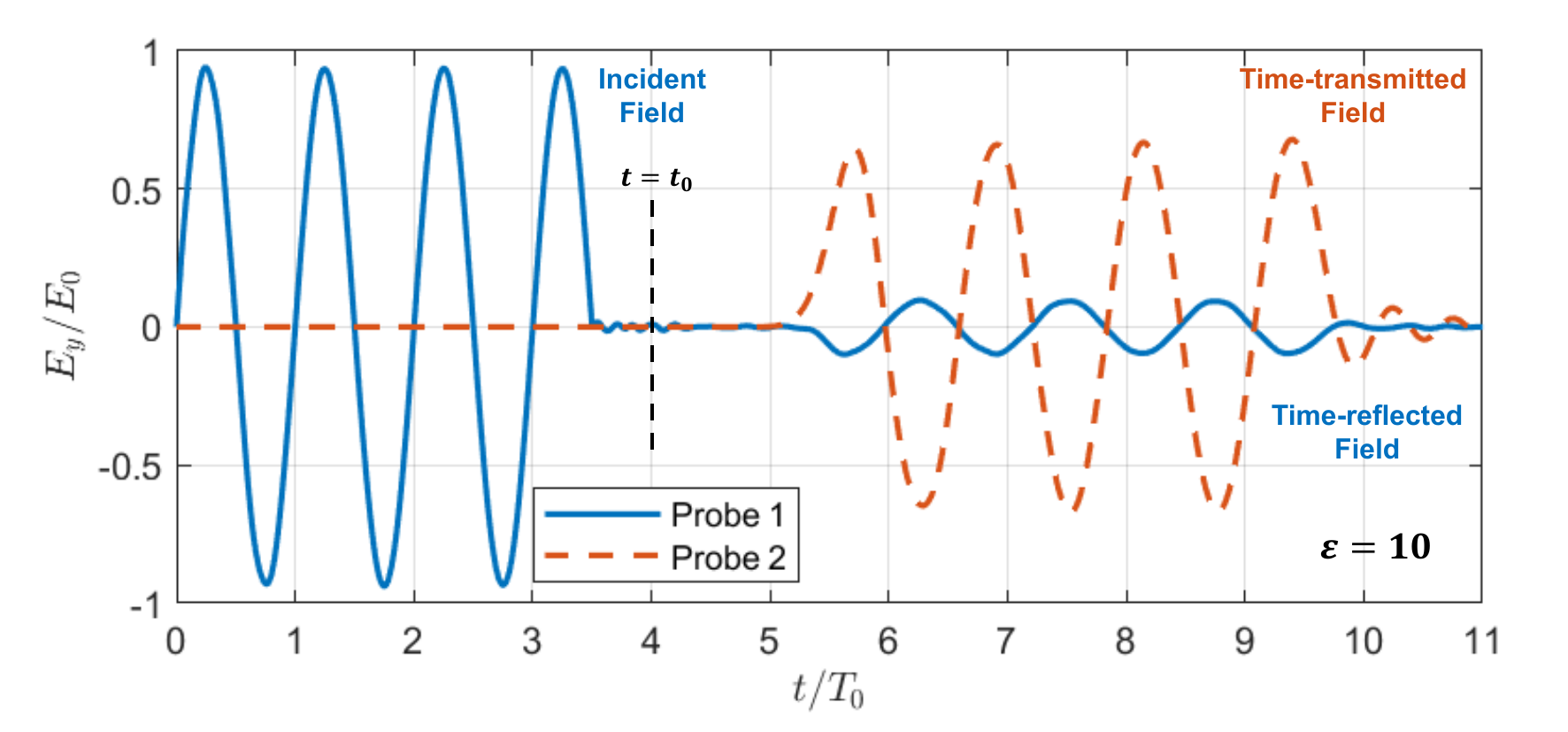}
            \put(0,40){(b)}
        \end{overpic}
    \end{subfigure}
    \hfill
    \begin{subfigure}{0.48\textwidth}
        \begin{overpic}[width=\linewidth]{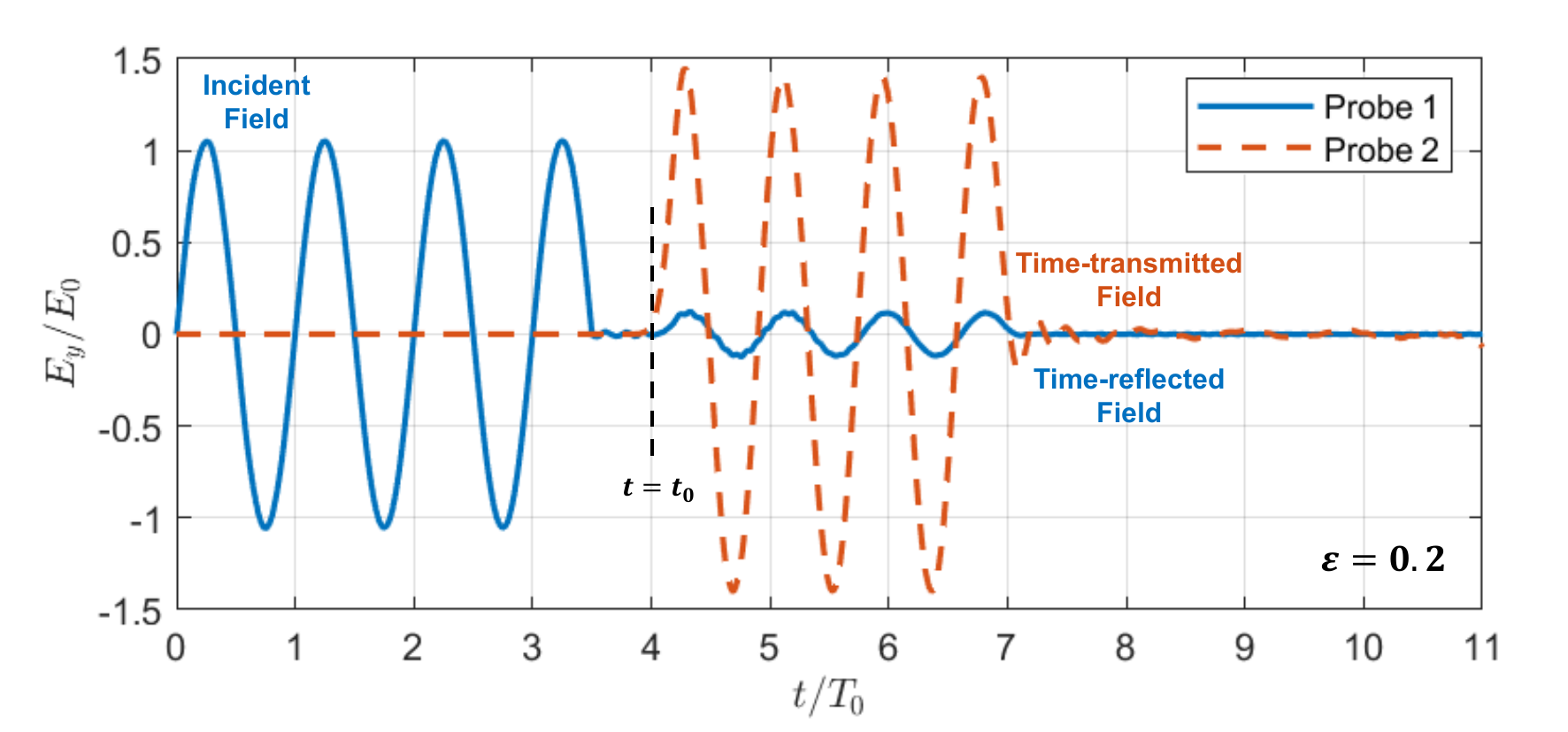}
            \put(0,40){(c)}
        \end{overpic}
    \end{subfigure}
   \caption{Full-wave time-domain simulations of an isotropic temporal interface. (a) Computational geometry comprising 50 cylinders along $x$, with periodic boundary conditions (PBCs) along $y$. The cylinder radii change simultaneously from $r_1$ to $r_2$ at $t=t_0$. (b,c) Normalized probe signals for $\varepsilon=10$ and $\varepsilon=0.2$, showing attenuation and amplification of the transmitted pulse, respectively. Solid blue curves record the incident and time-reflected pulses at Probe~1. Dashed orange curves record the time-transmitted pulse at Probe~2. Fields and time are normalized to the excitation amplitude $E_0$ and carrier period $T_0$ (before the temporal interface), respectively. Vertical dashed lines mark the switching instant.}
    \label{fig:isotropic_results}
\end{figure*}

Next, we test the effective-medium description using full-wave time-domain simulations in COMSOL Multiphysics for the actual array of cylinders undergoing an abrupt radius change. As illustrated in Fig.~\ref{fig:isotropic_results}(a), the two-dimensional computational domain contains 50 cylinders along the propagation direction ($x$), spanning a length $L=50a$, and one lattice period along $y$. Periodic boundary conditions along $y$ represent an array of infinite transverse extent, while scattering boundary conditions at the left and right boundaries allow outgoing waves to leave the domain while avoiding spurious reflections. Two field probes near these boundaries record the electric-field component $E_y(t)$.

All lengths are normalized to the incident wavelength $\lambda_1$ in the host medium. The lattice period is $a\approx0.091\lambda_1$ in both directions. The initial cylinder radius is chosen as $r_1\approx0.020\lambda_1$, ensuring an initial filling fraction of  $f_1=0.15$. The radius increases to $r_2=c r_1\approx0.033\lambda_1$, with $c\approx1.639$, giving a final filling fraction $f_2=c^2f_1\approx0.403$.

The array is excited from the left by a $y$-polarized pulse containing 3.5 carrier cycles at the wavelength of $\lambda_1$. All cylinder radii switch simultaneously at $t=t_0$, chosen such that the incident pulse is approximately centered within the array. Figures~\ref{fig:isotropic_results}(b) and (c) show the probe signals for permittivity contrasts $\varepsilon=10$ and $\varepsilon=0.2$, respectively. At Probe~1, the solid blue curve records the incident pulse followed by the backward-propagating, time-reflected pulse. At Probe~2, the dashed curve records the forward-propagating, time-transmitted pulse. The reflected and transmitted peak amplitudes are normalized to the incident peak amplitude for comparison with the analytical scattering coefficients.

For $\varepsilon=10$, the increase in effective permittivity produces attenuation and a frequency downshift, as seen in Fig.~\ref{fig:isotropic_results}(b). The simulated and theoretical temporal transmission coefficients are $|T_{\mathrm{sim}}|\approx0.713$ and $|T_{\mathrm{theory}}|\approx0.724$, while the corresponding temporal reflection coefficients are $|R_{\mathrm{sim}}|\approx0.100$ and $|R_{\mathrm{theory}}|\approx0.080$. For $\varepsilon=0.2$, the effective permittivity decreases, producing amplification and a frequency upshift [Fig.~\ref{fig:isotropic_results}(c)]. In this case, $|T_{\mathrm{sim}}|\approx1.330$ and $|T_{\mathrm{theory}}|\approx1.305$, whereas $|R_{\mathrm{sim}}|\approx0.110$ and $|R_{\mathrm{theory}}|\approx0.114$.

Thus, the transmission coefficients agree within $2\%$ in both cases. The reflection coefficients also agree closely for $\varepsilon=0.2$, while for $\varepsilon=10$ their absolute difference of $0.020$ corresponds to a larger relative discrepancy of $25\%$. These differences may arise from limitations of the Maxwell--Garnett approximation (small but finite array period compared to the wavelength) and from residual reflections at the computational boundaries.

\section{Isotropic-to-Anisotropic Temporal Interface}
\label{anisotropic}

We now consider an area-preserving temporal transformation from circular to elliptical cylinders, as shown in Fig.~\ref{fig:geometry}(b). At $t=t_0$, each cylinder of radius $r_1$ acquires semi-axes $a_2$ and $b_2$ along $x$ and $y$, respectively, with $a_2b_2=r_1^2$. The filling fraction therefore remains unchanged, while the unequal depolarization factors introduce an anisotropic effective response. Before the transition, the in-plane effective permittivity is the scalar $\varepsilon_{\mathrm{eff},1}$. After it, the permittivity tensor has principal components $\varepsilon_{\mathrm{eff},2x}$ and $\varepsilon_{\mathrm{eff},2y}$, as defined in Eq.~\eqref{eq1new}. All effective permittivities are assumed real, positive, and frequency independent, and the permeability remains $\mu_0$.

\begin{figure*}[tb]
    \centering
    \includegraphics[width=0.8\textwidth]{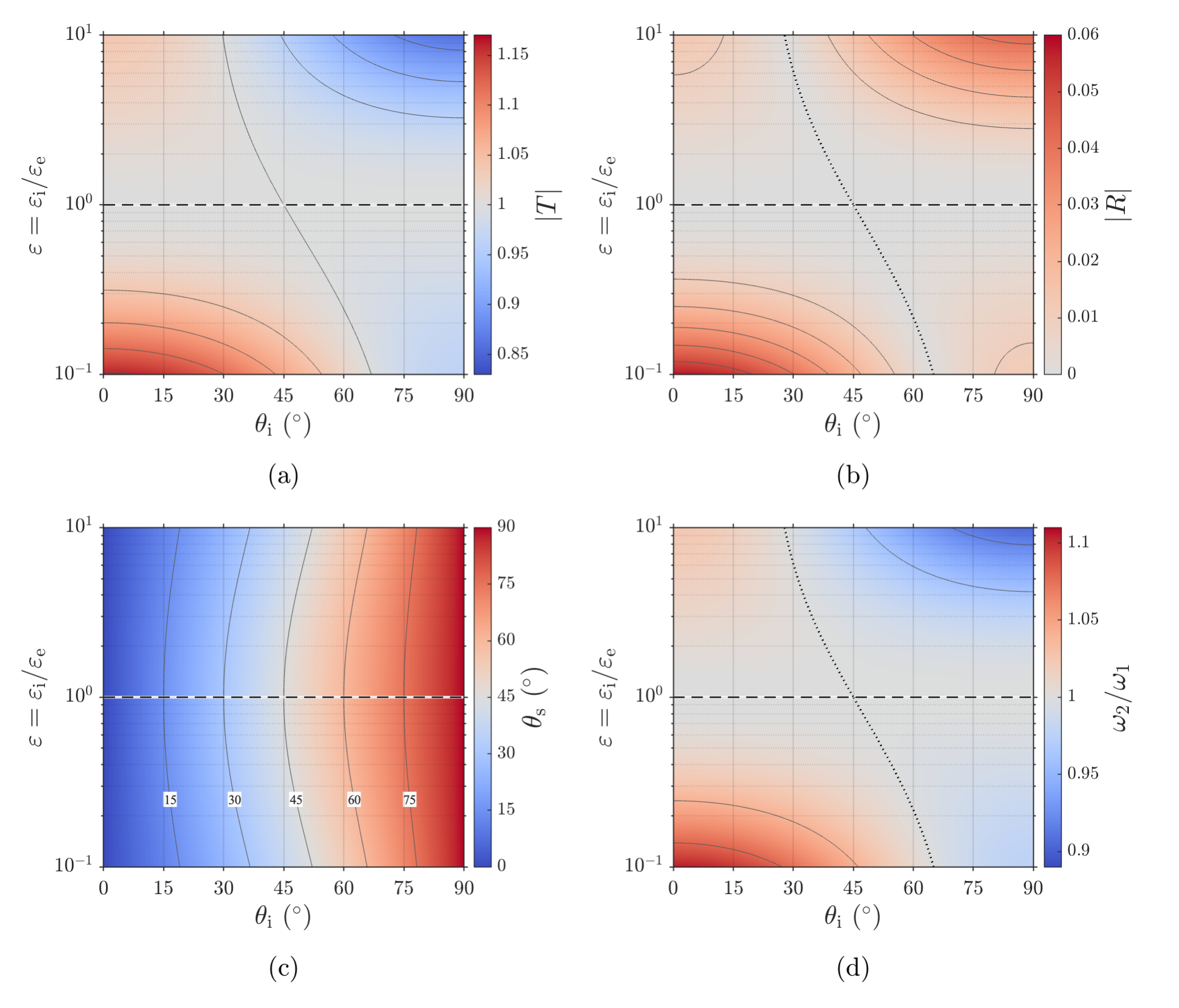}
    \caption{Analytical response of an isotropic-to-anisotropic temporal interface versus the incident-wavevector angle $\theta_{\mathrm{i}}$ and permittivity contrast $\varepsilon=\varepsilon_{\mathrm{i}}/\varepsilon_{\mathrm{e}}$, for $f_1=f_2=0.08$, $a_2=3r_1$, and $b_2=r_1/3$. (a) Transmission amplitude $|T|$. (b) Reflection amplitude $|R|$. (c) Transmitted energy-flow angle $\theta_{\mathrm{s}}$. (d) Frequency ratio $\omega_2/\omega_1$. Horizontal dashed lines mark $\varepsilon=1$. Dotted curves in (b,d) indicate reflection-free directions, where $\omega_2=\omega_1$. Thin solid curves are contours of the plotted quantities.}
    \label{fig:pseudocolorplot_anisotropic}
\end{figure*}

The incident TM-polarized plane wave has its magnetic field along $z$ and propagates in the $xy$-plane with wavevector
\begin{align}
    \mathbf{k}
    =k\bigl(\cos\theta_{\mathrm{i}}\,\hat{\mathbf{x}}
           +\sin\theta_{\mathrm{i}}\,\hat{\mathbf{y}}\bigr).
    \label{eq:incident_wavevector_aniso}
\end{align}
Here, $\theta_{\mathrm{i}}$ is the angle between the incident wavevector and the positive $x$-axis. The incident electric field is directed along $-\sin\theta_{\mathrm{i}}\,\hat{\mathbf{x}}+\cos\theta_{\mathrm{i}}\,\hat{\mathbf{y}}$. Thus, $\theta_{\mathrm{i}}=0^\circ$ and $90^\circ$ describe propagation along the two principal axes. Within the homogeneous effective-medium description, the temporal modulation is spatially uniform and conserves both components of $\mathbf{k}$.

The dispersion relation after the transition is
\begin{align}
    \omega_2^2
    =\frac{1}{\mu_0}
    \left(
    \frac{k_x^2}{\varepsilon_{\mathrm{eff},2y}}
    +\frac{k_y^2}{\varepsilon_{\mathrm{eff},2x}}
    \right),
    \label{eq:dispersion_aniso}
\end{align}
where the effective permittivities denote absolute permittivities. Combining this relation with the incident dispersion relation gives the angle-dependent frequency conversion~\cite{pacheco2020temporal},
\begin{align}
    \frac{\omega_2}{\omega_1}
    =\sqrt{
    \frac{\varepsilon_{\mathrm{eff},1}}
         {\varepsilon_{\mathrm{eff},2x}}
    \sin^2\theta_{\mathrm{i}}
    +\frac{\varepsilon_{\mathrm{eff},1}}
          {\varepsilon_{\mathrm{eff},2y}}
    \cos^2\theta_{\mathrm{i}}}.
    \label{eq:freq_ratio_aniso}
\end{align}
Here, $\omega_1$ and $\omega_2$ are positive. At the conserved wavevector, the transmitted and reflected branches have frequencies $+\omega_2$ and $-\omega_2$, respectively.

To express the temporal scattering coefficients, we define the dimensionless factor
\begin{align}
    \mathcal{A}(\theta_{\mathrm{i}})
    =\sqrt{
    \left(\frac{\varepsilon_{\mathrm{eff},1}}
               {\varepsilon_{\mathrm{eff},2x}}\right)^2
    \sin^2\theta_{\mathrm{i}}
    +\left(\frac{\varepsilon_{\mathrm{eff},1}}
                {\varepsilon_{\mathrm{eff},2y}}\right)^2
    \cos^2\theta_{\mathrm{i}}}.
    \label{eff_vector_norm1}
\end{align}
Continuity of $\mathbf{D}$ and $\mathbf{B}$ at $t=t_0$ then gives~\cite{pacheco2020temporal}
\begin{align}
    T&=\frac{\mathcal{A}}{2}
    \left(1+\frac{\omega_1}{\omega_2}\right),
    \label{eff_T_aniso}\\
    R&=\frac{\mathcal{A}}{2}
    \left(1-\frac{\omega_1}{\omega_2}\right).
    \label{eff_R_aniso}
\end{align}
These coefficients normalize the outgoing electric-field amplitudes to the magnitude of the incident field, with phases referenced to $t=t_0$.

The introduced anisotropy also redirects the energy flow, producing temporal aiming~\cite{pacheco2020temporal}. Although $\mathbf{k}$ is conserved, the transmitted Poynting vector is directed along $\bigl(k_x/\varepsilon_{\mathrm{eff},2y},k_y/\varepsilon_{\mathrm{eff},2x}\bigr)$. Its angle $\theta_{\mathrm{s}}$, measured from the positive $x$-axis, therefore satisfies
\begin{align}
    \theta_{\mathrm{s}}
    =\arctan\!\left(
    \frac{\varepsilon_{\mathrm{eff},2y}}
         {\varepsilon_{\mathrm{eff},2x}}
    \tan\theta_{\mathrm{i}}\right),
    \label{eq:eff_poynting_angle}
\end{align}
for $0\leq\theta_{\mathrm{i}}\leq\pi/2$. The angular deflection is $\theta_{\mathrm{s}}-\theta_{\mathrm{i}}$, and the reflected branch carries energy in the opposite direction to the transmitted branch. Along either principal axis, or when the final medium is isotropic, $\theta_{\mathrm{s}}=\theta_{\mathrm{i}}$ and no deflection occurs.

Figure~\ref{fig:pseudocolorplot_anisotropic} summarizes the angular dependence of temporal scattering, energy-flow direction, and frequency conversion for $0.1\leq\varepsilon\leq10$. We use the area-preserving deformation $a_2=3r_1$ and $b_2=r_1/3$, giving an aspect ratio $a_2/b_2=9$ and depolarization factors $N_{2x}=0.1$ and $N_{2y}=0.9$. The filling fraction is fixed at $f_1=f_2=0.08$. For this aspect ratio, nonoverlapping inclusions on a square lattice require $f\leq\pi/36\approx0.0873$, which is satisfied by the chosen geometry. Along the dashed line at $\varepsilon=1$, the inclusions and host are electromagnetically identical, giving $|T|=1$, $|R|=0$, $\omega_2/\omega_1=1$, and $\theta_{\mathrm{s}}=\theta_{\mathrm{i}}$.

The angular dependence in Fig.~\ref{fig:pseudocolorplot_anisotropic}(a) follows from the changes in the principal permittivity components. Elongating the inclusions along $x$ while conserving their cross-sectional area decreases $N_x$ and increases $N_y$ from their initial value of $1/2$. According to Eq.~\eqref{eq:MG_elliptical_cylinders}, decreasing the depolarization factor at a fixed filling fraction increases the corresponding effective permittivity for both $\varepsilon>1$ and $\varepsilon<1$. The deformation therefore raises the $x$-component and lowers the $y$-component relative to the initial isotropic permittivity.
For propagation along $x$ ($\theta_{\mathrm{i}}=0^\circ$), the electric field is directed along $y$ and experiences the decrease in the $y$-component of permittivity, producing transmitted-field enhancement. Conversely, for propagation along $y$ ($\theta_{\mathrm{i}}=90^\circ$), the electric field is directed along $x$ and experiences an increase in permittivity, producing attenuation.  

The reflected amplitude in Fig.~\ref{fig:pseudocolorplot_anisotropic}(b) remains small, with a maximum of approximately $0.0584$. In addition to vanishing at $\varepsilon=1$, reflection disappears along the dotted curve where $\omega_2/\omega_1=1$, as follows from Eq.~\eqref{eff_R_aniso}. This condition defines the temporal Brewster direction. For example, it occurs at $\theta_{\mathrm{i}}\approx27.8^\circ$ for $\varepsilon=10$ and $\theta_{\mathrm{i}}\approx65.1^\circ$ for $\varepsilon=0.1$. The transmitted electric-field amplitude can exceed unity at these angles despite the absence of frequency conversion, because the temporal transition changes the field polarization.

Figure~\ref{fig:pseudocolorplot_anisotropic}(c) shows the transmitted energy-flow angle. Since $\varepsilon_{\mathrm{eff},2y}/\varepsilon_{\mathrm{eff},2x}<1$, the energy flow bends toward the $x$-axis for all oblique directions in both contrast regimes. The deflection vanishes at $0^\circ$ and $90^\circ$ and becomes larger as the contrast departs from unity. For example, at $\theta_{\mathrm{i}}=45^\circ$, both $\varepsilon=0.1$ and $\varepsilon=10$ give $\theta_{\mathrm{s}}\approx37.9^\circ$, corresponding to a deflection of approximately $7.1^\circ$ toward $x$. 

Finally, Fig.~\ref{fig:pseudocolorplot_anisotropic}(d) shows that propagation near the $x$-axis produces a frequency upshift, while propagation near the $y$-axis produces a downshift. The dotted reflection-free curve separates these regimes.  Thus, the same area-preserving shape change controls field amplitude, frequency, and energy-flow direction through the incident propagation angle.

\begin{figure*}[t]
    \centering

    % Left column
    \begin{minipage}{0.47\textwidth}
        \centering

        % (a) - portrait, spanning two rows
        \begin{subfigure}{0.8\linewidth}
            % \makebox[0pt][r]{(a)\hspace{0.5em}}%
           \begin{overpic}[width=\linewidth]{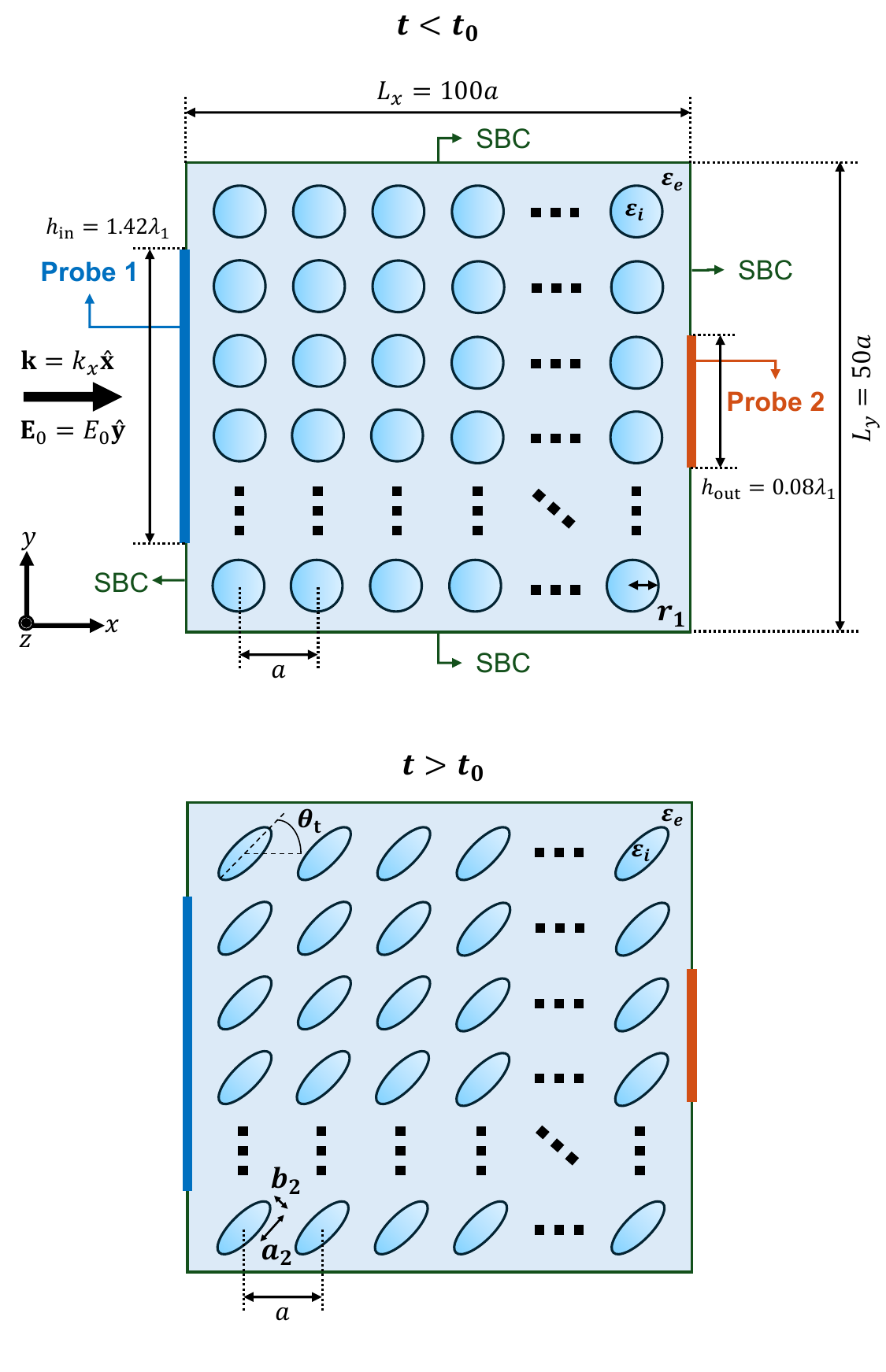}
            \put(-4,95){(a)}
        \end{overpic}
            \label{fig:a}
        \end{subfigure}

        % (d) - third row
        \begin{subfigure}{0.9\linewidth}
            % \makebox[0pt][r]{(d)\hspace{0.5em}}%
             \begin{overpic}[width=\linewidth]{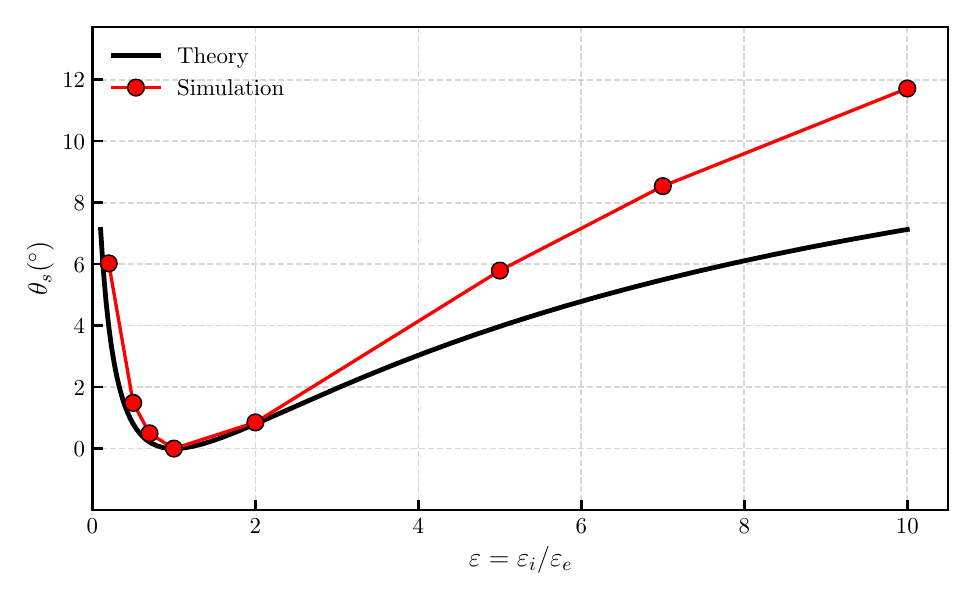}
            \put(0,60){(d)}
        \end{overpic}
            \label{fig:d}
        \end{subfigure}

    \end{minipage}
    \hfill
    % Right column
    \begin{minipage}{0.5\textwidth}
        \centering

        % (b)
        \begin{subfigure}{0.85\linewidth}
            % \makebox[0pt][r]{(b)\hspace{0.5em}}%
             \begin{overpic}[width=\linewidth]{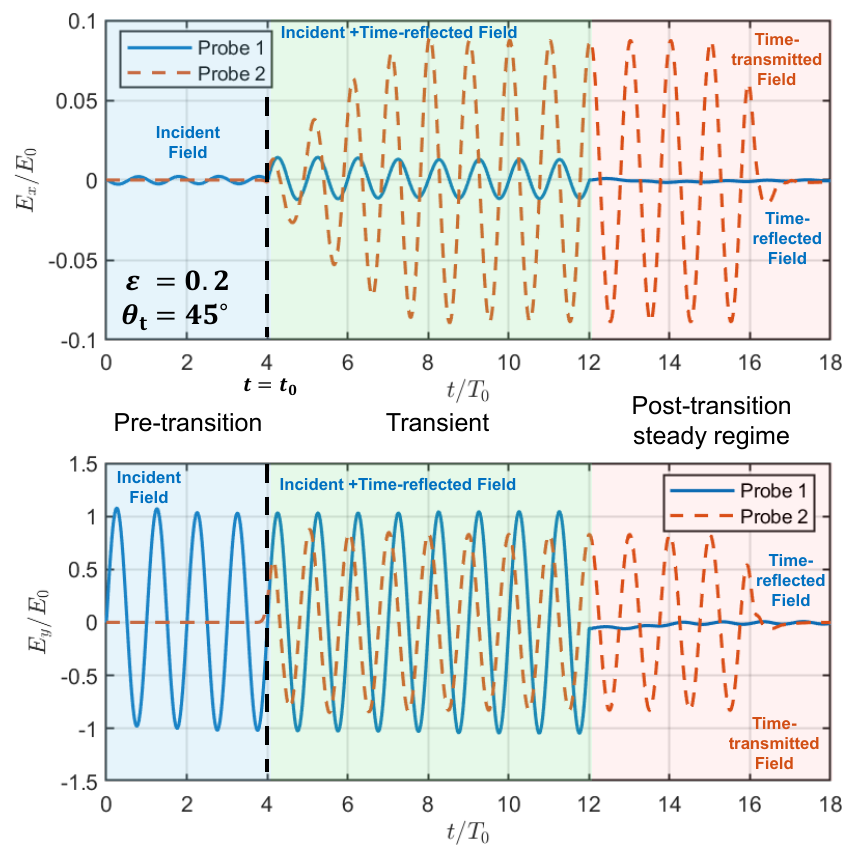}
            \put(-3,95){(b)}
        \end{overpic}
            \label{fig:b}
        \end{subfigure}

        % (c)
        \begin{subfigure}{0.85\linewidth}
            % \makebox[0pt][r]{(c)\hspace{0.5em}}%
             \begin{overpic}[width=\linewidth]{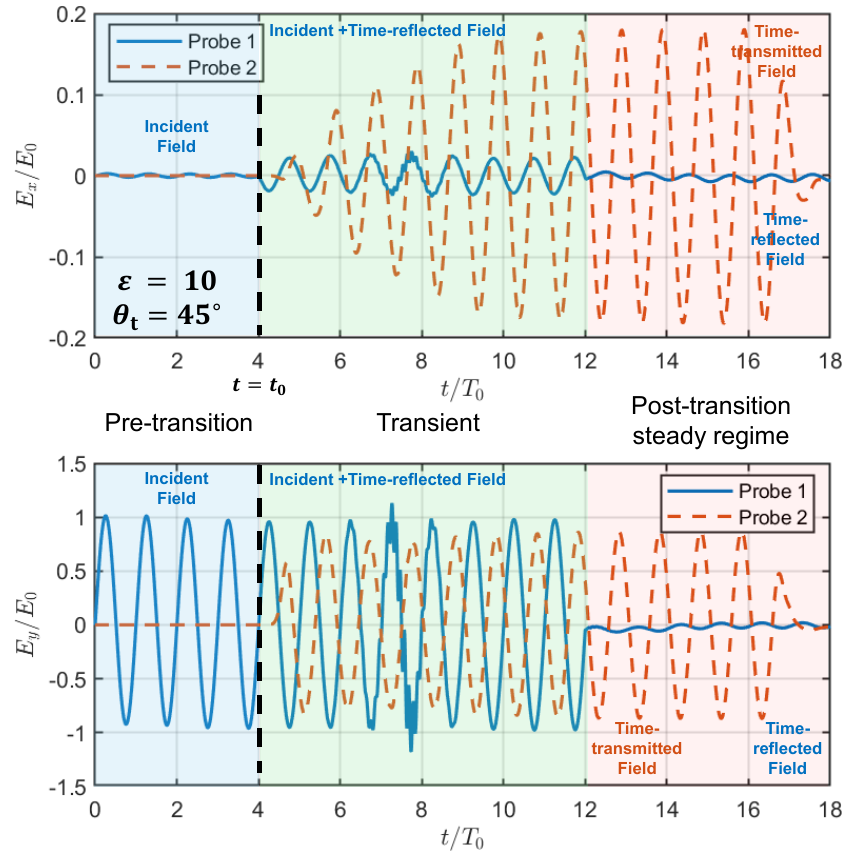}
            \put(-3,95){(c)}
        \end{overpic}
            \label{fig:c}
        \end{subfigure}

    \end{minipage}

\caption{Full-wave simulations of temporal aiming. (a) Simulation geometry before and after the temporal transition. The incident wave propagates along $x$, while the elliptical inclusions have their major axes tilted by $\theta_{\rm t}=45^\circ$. (b,c) Normalized electric-field components recorded at Probe 1 (solid blue) and Probe 2 (dashed orange) for $\varepsilon=0.2$ and $\varepsilon=10$, respectively. The temporal response is divided into three regimes: the pre-transition regime ($t<t_0$), the transient regime immediately after the temporal transition, and the post-transition steady regime. The vertical dashed lines mark the switching instant $t=t_0$. (d) Transmitted Poynting-vector angle $\theta_{\rm s}$, measured from the laboratory $x$-axis, versus permittivity contrast $\varepsilon=\varepsilon_i/\varepsilon_e$: effective-medium theory (black curve) and full-wave simulations (red markers).}
    \label{fig:anisotropic_aiming}
\end{figure*}

We next assess the effective-medium predictions using full-wave time-domain simulations in COMSOL Multiphysics. Figure~\ref{fig:anisotropic_aiming}(a) shows the two-dimensional simulation geometry, consisting of a square-lattice array with $100$ cylinders along $x$ and $50$ cylinders along $y$, corresponding to dimensions $L_x=100a$ and $L_y=50a$. Scattering boundary conditions (SBCs) are applied at the outer boundaries to allow outgoing waves to leave the computational domain. The incident wave propagates along the laboratory $x$-axis, with its electric and magnetic fields directed along $y$ and $z$, respectively. Finite-length Probe~1 and Probe~2, located near the left and right boundaries, record the electric-field components.

The orientation of the anisotropy axes in these simulations differs from that used above (see e.g.,  Eq.~(\ref{eq:freq_ratio_aniso})). There, the principal axes coincide with the coordinate axes $x$ and $y$, and the incident wavevector makes an angle $\theta_{\rm i}$ with the $x$-axis. Here, the wave is normally incident on the left boundary and propagates along $+x$, while the principal axes of the anisotropic medium are tilted. Specifically, at $t=t_0$, the circular cylinders transform into elliptical cylinders whose major axes make an angle $\theta_{\rm t}$ with the laboratory $x$-axis, as indicated in Fig.~\ref{fig:anisotropic_aiming}(a). We choose $\theta_{\rm t}=\theta_{\rm i}=45^\circ$ to compare configurations with the same magnitude of the angle between the incident wavevector and the major principal axis. The deformation is otherwise unchanged: $a_2=3r_1$ and $b_2=r_1/3$, preserving the filling fraction at $f=0.08$.
The square-lattice period is
$a\approx0.0417\lambda_1$, and the initial cylinder radius is
$r_1=0.16a\approx0.0067\lambda_1$.
The structure is illuminated by a light pulse containing 12 carrier cycles at the wavelength of $\lambda_1$.

Figures~\ref{fig:anisotropic_aiming}(b) and (c) show the
simulated electric-field components for $\varepsilon=0.2$
and $\varepsilon=10$, respectively, with $E_x$ in the upper
plots and $E_y$ in the lower plots. The fields are normalized
to the incident amplitude $E_0$, and time is normalized to
the incident period $T_0$, with switching at $t_0=4T_0$.
The solid blue traces at Probe~1 contain the incident and
returning time-reflected contributions, whereas the dashed
orange traces at Probe~2 record the time-transmitted field.
Before switching, the incident field is 
$y$-polarized, with a negligible $E_x$ component due to numerical noise. The temporal
transition generates a nonzero $E_x$ component, while $E_y$
remains dominant. This change in polarization is associated
with the redirection of energy flow enabled by the tilted
anisotropy.

To quantify this redirection, we define $\theta_{\rm s}$
as the angle of the transmitted Poynting vector measured
from the positive laboratory $x$-axis. Since the incident
wave propagates along this axis, $\theta_{\rm s}$ also
directly measures the deflection from its initial direction.
For the counterclockwise tilt considered here, the incident
wavevector makes a signed angle $-\theta_{\rm t}$ with the
major principal axis. Transforming
Eq.~\eqref{eq:eff_poynting_angle} into the laboratory
coordinates therefore gives
\begin{align}
    \theta_{\rm s}
    &=
    \theta_{\rm t}
    -
    \arctan\left(
    \frac{\varepsilon_{\mathrm{eff},2b}}
         {\varepsilon_{\mathrm{eff},2a}}
    \tan\theta_{\rm t}
    \right),
    \label{eq:eff_poynting_angle_tilted}
\end{align}
where $\varepsilon_{\mathrm{eff},2a}$ and
$\varepsilon_{\mathrm{eff},2b}$ are the principal effective
permittivities along the major and minor ellipse axes,
respectively. They correspond to
$\varepsilon_{\mathrm{eff},2x}$ and
$\varepsilon_{\mathrm{eff},2y}$ in the unrotated geometry.

The simulated angles are extracted from the polarization
of the time-transmitted field recorded at Probe~2. For each simulation, the steady-state interval was identified by inspection of the time-domain probe signals. The positive peak amplitudes of $E_x$ and $E_y$ within this interval were then identified for each cycle, and their average values were used as $E_x$ and $E_y$, respectively. For the linearly polarized TM wave, the Poynting vector
is perpendicular to the electric field. The positive
deflection angle considered here is therefore obtained as
\begin{align}
    \theta_{\rm s}
    =
    \arctan\left(\frac{E_x}{E_y}\right),
\end{align}
where $E_x$ and $E_y$ are the averaged positive amplitudes of the
transmitted electric-field components, extracted over
the same time window.

Figure~\ref{fig:anisotropic_aiming}(d) compares these
simulated angles with the analytical predictions of
Eq.~\eqref{eq:eff_poynting_angle_tilted} as the permittivity
contrast is varied. Both approaches predict zero deflection
at $\varepsilon=1$, where the inclusions are
electromagnetically indistinguishable from the host.
The deflection increases as the contrast departs from unity
in either direction. The simulations reproduce this overall
trend but predict larger angles at stronger contrasts.
These discrepancy at higher contrasts indicate quantitative limitations of
the quasistatic Maxwell--Garnett description for the
simulated structure. 
For $\varepsilon\gg1$, increasing the permittivity contrast
at fixed geometry and incident wavelength shortens the
wavelength inside the cylinders, bringing their response
closer to Mie resonances. Therefore, the associated dynamic effects increases deviations from the quasistatic
Maxwell--Garnett prediction.

Nevertheless, the larger deflections obtained in the full-wave simulations
motivate further investigation of temporal wave control
beyond the quasistatic regime. In particular, engineering
the structural resonances of the inclusions may provide
additional control over temporal scattering and aiming.
Resonance-enhanced temporal reflection has recently been
demonstrated theoretically~\cite{li2026resonance}. Extending
this approach to the present anisotropic geometry would
require a dispersive treatment of the temporal transition.

\section*{Conclusions}

We have investigated temporal scattering in metamaterials whose cylindrical inclusions undergo abrupt changes in size or cross-sectional shape. Combining quasistatic homogenization with temporal boundary conditions, we showed that radius modulation controls the scalar effective permittivity, enabling temporal reflection, amplification or attenuation, and frequency conversion. An area-preserving transformation from circular to elliptical cylinders introduces anisotropy, producing angle-dependent scattering and redirecting the energy flow while conserving the wavevector. Full-wave time-domain simulations reproduce the main theoretical trends and predict larger aiming angles at higher permittivity contrasts, motivating further investigation beyond the quasistatic approximation.

These findings represent an initial step toward the practical realization of anisotropic temporal interfaces through geometric modulation. Laser-induced plasma offers a particularly promising platform: spatially patterned femtosecond excitation can generate plasma filament arrays~\cite{gao2013femtosecond}, and ultrafast plasma time boundaries have already enabled experimental frequency conversion of terahertz waves~\cite{huang2025terahertz}. Realizing the proposed shape transformations will require synchronized control of the plasma geometry and an extension of the present theory to include plasma dispersion, collisional losses, and finite formation and relaxation times. Understanding these dynamics will also be essential for implementing repeated temporal switching.

The approach could be extended to suitably designed chiral and other bianisotropic meta-atoms, allowing geometric modulation to control magnetoelectric coupling at temporal interfaces~\cite{mostafa2023spin,Mirmoosa2024}. Natural chiral materials typically exhibit weak chiroptical responses, whereas engineered metamaterials can support much stronger responses governed by their geometry~\cite{kilic2024controlling}. Shape variation could therefore enable substantial temporal changes in magnetoelectric coupling and enhance polarization-dependent frequency conversion and amplification. With suitable periodic switching protocols, these possibilities could be extended to chiral and bianisotropic photonic time crystals, offering control over polarization-selective momentum gaps and parametric amplification~\cite{koufidis2024electromagnetic}.

\bibliography{apssamp}

\end{document}